\documentclass[aps, superscriptaddress, showpacs]{revtex4}
\usepackage{graphicx}
\usepackage{amsmath}
\usepackage{bm}
\usepackage{amssymb}

\usepackage[pdftex,bookmarks=false,hypertexnames=false,breaklinks=true,colorlinks=true,urlcolor=blue,citecolor=blue,linkcolor=blue]{hyperref}

\newcommand{\be}{\begin{equation}}
\newcommand{\ee}{\end{equation}}
\newcommand{\ba}{\begin{eqnarray}}
\newcommand{\ea}{\end{eqnarray}}

\begin{document}

\title{Screening of a charged impurity in graphene in a magnetic field}
\date{\today}

\author{O. O. Sobol}
\affiliation{Department of Physics, Taras Shevchenko National University of Kiev, Kiev, 03680, Ukraine}

\author{P. K. Pyatkovskiy}
\affiliation{Department of Physics and Astronomy, University of Manitoba, Winnipeg, R3T 2N2, Canada}

\author{E. V. Gorbar}
\affiliation{Department of Physics, Taras Shevchenko National University of Kiev, Kiev, 03680, Ukraine}
\affiliation{Bogolyubov Institute for Theoretical Physics, Kiev, 03680, Ukraine}

\author{V. P. Gusynin}
\affiliation{Bogolyubov Institute for Theoretical Physics, Kiev, 03680, Ukraine}

\begin{abstract}
The electron states in the field of a charged impurity in graphene in a magnetic field are studied numerically. It is shown that a charged impurity removes the degeneracy of Landau levels converting
them into bandlike structures. As the charge of impurity grows, the repulsion of sublevels of different Landau levels with the same value of orbital momentum takes place leading to the redistribution of the
wave function profiles of these sublevels near the impurity. By studying the polarization effects, it
is shown in agreement with the recent experiments that the effective charge of impurity can be very effectively tuned by chemical potential. If the chemical potential is situated inside a Landau level,
then the charge of impurity is strongly diminished. In addition, the polarization function in this case
has a peak at zero momentum, which leads to the sign-changing oscillations of the screened potential as
a function of distance. If the chemical potential lies between the Landau levels, then the screened potential does not change sign, the screening is minimal, and the charged impurity can strongly affect
the electron spectrum.
\end{abstract}
\pacs{81.05.ue, 73.22.Pr}
\maketitle

\section{Introduction}

After the experimental discovery of graphene \cite{Novoselov} whose quasiparticles are described by a relativistic-like (2+1)-dimensional Dirac equation with the velocity of light $c$ replaced by the
Fermi velocity $v_F \approx c/300$, it was soon realized that this material is very promising for
the experimental observation of the atomic collapse \cite{Pomeranchuk,Zeldovich,Greiner} in the
Coulomb field of a point charge $Ze$. Indeed, the large value of the coupling constant $\alpha_g=e^2/
(\hbar v_F)\approx 2.2$ leads to a dramatically smaller value of the critical charge $Z_c \approx
1/2$ [\onlinecite{Pereira,Shytov,Novikov}] compared to that in quantum electrodynamics where
$Z_c\gtrsim 170$ [\onlinecite{Zeldovich,Greiner}]. Since nuclei with such large charges do not
exist in nature, this phenomenon was never observed in quantum electrodynamics. The supercritical
Coulomb center instability is also closely related to the excitonic instability in graphene in the
strong-coupling regime $\alpha_g>\alpha_c\sim1$ (see
Refs. [\onlinecite{excitonic-instability,Fertig,Guinea}]) and the gap opening, which may transform
graphene into an insulator \cite{metal-insulator,GGG2010,MS-phase-transition,Gonzalez}.
It is well known that the charged impurities are the dominant source of scatterers
in graphene affecting its major electron transport features (see, for example,
Ref. [\onlinecite{Fan}] and references therein).

Recently, by creating artificial nuclei in a certain region of graphene fabricated through the
deposition of charged calcium dimers on graphene with the tip of a scanning tunneling microscope,
the supercritical regime was reached and the resonances corresponding to the atomic collapse
states were observed \cite{Wang,states}. The supercritical instability for ${\rm C}a$ dimers on graphene
was theoretically studied in Ref. [\onlinecite{Kirczenow}] by making use of the density functional
theory and an improved Huckel model. In a recent publication, Mao \textit{et al}. proposed a more effective
way to deposit a charge in graphene: they showed that a single-atom vacancy can host a local charge
that can be gradually changed by applying the voltage pulses with the tip of a scanning tunneling
microscope \cite{Mao}. Similarly to the case of charged adatoms on graphene, a transition into a supercritical
 regime was observed with the formation of quasibound states at the vacancy site.

In the continuum model, three of us extended \cite{two-centers} the study of the supercritical instability
of a single Coulomb center in gapped graphene to the case of the simplest cluster of two equally charged impurities when the charges of impurities are subcritical, whereas their total charge exceeds a critical one. We determined the critical distance between the impurities separating the supercritical and
subcritical regimes as a function of charges of impurities and a gap.

An interesting problem of two oppositely charged impurities was considered in Refs.~[\onlinecite{Egger,Matrasulov}]. Obviously, this dipole problem is particle-hole symmetric and electron states are symmetric with respect to the change  of the sign of energy, $E \to -E$. Naively,
one would think that the supercritical regime in this problem sets in when the lowest-energy electron
bound level intersects the highest energy hole bound state at $E=0$. However, it was found \cite{Egger} that  the levels first approach each other as the dipole moment increases, and then diverge. In fact,
this behavior is typical for an avoided crossing \cite{Wigner} of the states with the same symmetry.
The issue of supercriticality was revisited by three of us in Ref.~[\onlinecite{migration}]. We showed that
a new type of supercritical behavior is realized in the dipole problem, which is connected with the
change of localization of the  highest-energy occupied state  from the negatively charged impurity to
the positively charged one. Such a migration of the wave function corresponds to an electron and a hole spontaneously created from the vacuum in bound states and screening the positively and negatively charged impurities of the supercritical electric dipole, respectively.

Magnetic fields and their effects are ubiquitous in physics. Electron states in a magnetic field are described by the infinitely degenerate Landau levels. Since all electron states of a given Landau level
have the same energy, magnetic field completely quenches the kinetic energy of these electrons making
such systems ideally suitable for the realization of interaction-driven phases of matter. It is worth
recalling here just two notable examples such as the fractional quantum Hall effect \cite{fractional}
and magnetic catalysis \cite{magnetic-catalysis}. In view of the above, it is
interesting how magnetic field affects the atomic collapse in graphene \cite{magnetic-instability,Barlas,Kim-Yang}.
The crucial ingredient  is the existence of an infinitely degenerate zero-energy Landau
level for gapless Dirac fermions in a magnetic field. In this case, any small attractive potential
leads to the appearance of negative-energy bound states, in contrast to the case without magnetic field
when the impurity charge must exceed a certain critical value for the appearance of quasibound states.
Let us mention also that this result presents a quantum-mechanical single-particle counterpart of the magnetic catalysis in graphene. Recently, electron states in the field of several charged impurities in graphene in a magnetic field were considered in Ref.~[\onlinecite{Slizovskiy}].

Experimentally, electron states in the field of a Coulomb impurity in a magnetic field were studied
in Refs.~[\onlinecite{Mao,states}], where it was shown that the strength of a charged impurity can be tuned
by controlling the occupation of Landau-level states with a gate voltage. At low Landau level occupation, the screening is so effective that the impurity becomes practically invisible, whereas at full occupancy
the screening is weak and the impurity attains its maximum strength. In this regime, the first experimental observation of Landau-level splitting into discrete states due to lifting the orbital degeneracy was reported. This experiment provides a motivation for the present study where, in contrast to the earlier theoretical treatments \cite{magnetic-instability,Barlas,Kim-Yang,Sun-Zhu,Zhu2014,Slizovskiy}, the main accent is made on the role of polarization effects.

The paper is organized as follows. In Sec.~\ref{section-model}, we consider the electron levels in the
Coulomb field of a charged impurity in graphene in a magnetic field. How the screening of a charged impurity can be tuned by  chemical potential is studied in Sec.~\ref{section-screening}. The results are
summarized and discussed in Sec.~\ref{section-conclusion}.

\section{Charged impurity in a magnetic field}
\label{section-model}

Let us consider the electron states in graphene with a single charged impurity in a magnetic field. The
Dirac Hamiltonian in $2+1$ dimensions which describes the quasiparticle states in the vicinity of the $K_{\pm}$ points of graphene in the field of charged impurity in a magnetic field reads (although,
in view of the magnetic catalysis \cite{magnetic-catalysis}, a nonzero gap is always generated in
graphene in a perpendicular magnetic field \cite{metal-insulator}, this gap is rather small for
realistic magnetic fields; therefore, for simplicity, we neglect it in our analysis below)
\begin{equation}
H=v_{F}\boldsymbol{\sigma}\boldsymbol{\pi}+V(\mathbf{r}),
\label{general-Hamiltonian}
\end{equation}
where $\boldsymbol{\pi}=-i\hbar\boldsymbol{\nabla}+\frac{e}{c}\mathbf{A}$, $-e<0$ is the electron charge,
the vector potential $\mathbf{A}=B/2(-y,\ x)$ in the symmetric gauge describes magnetic field perpendicular to the plane of graphene, and $\boldsymbol{\sigma}$ are the Pauli
matrices. The Hamiltonian (\ref{general-Hamiltonian}) acts on a two-component spinor $\Psi_{\xi s}$ which
carries the valley ($\xi=\pm)$ and spin ($s=\pm$) indices and we use the standard convention: $\Psi^{T}_{+s}=(\psi_{A},\psi_{B})_{K_{+}s}$, whereas $\Psi^{T}_{-s}=(\psi_{B},-\psi_{A})_{K_{-}s}$,
and $A,B$ refer to two sublattices of the hexagonal graphene lattice. We regularize the Coulomb potential
of an impurity by introducing a parameter $a$ of the order of the graphene lattice spacing. Then the regularized interaction potential of the impurity with charge $Q=Ze$ is given by
\begin{equation}
V\left(\mathbf{r}\right)=-\frac{Ze^2}{\kappa\sqrt{r^{2}+a^{2}}},
\label{impurity-potential}
\end{equation}
where $\kappa$ is the dielectric constant. Since the interaction potential (\ref{impurity-potential})
does not depend on valley and spin, we will omit the valley and spin indices $\xi$ and $s$ in the wave
functions below.

It is convenient to introduce the magnetic length $l_{B}=\sqrt{{\hbar c}/{|eB|}}$
and the dimensionless quantity $\zeta=Z e^{2}/(\hbar v_{F})$ which characterizes the strength of the bare impurity. Since the total angular momentum is conserved, we use the polar coordinates $(r,\ \theta)$ to write
\begin{equation}
\Psi=\frac{1}{r}\left(\begin{array}{c}e^{-i(m+1)\theta}f(r)\\-i e^{-i m\theta}g(r)
\end{array}\right),
\end{equation}
where $m$ is the orbital quantum number. Then the Dirac equation takes the form
\begin{equation}
\label{system-eq}
\left\{
\begin{array}{c}
f'+\frac{m}{r}f-\frac{r}{2l_{B}^{2}}f-\frac{E-V(r)}{\hbar v_{F}}g=0,\\
g'-\frac{m+1}{r}g+\frac{r}{2l_{B}^{2}}g+\frac{E-V(r)}{\hbar v_{F}}f=0.
\end{array}
\right.
\end{equation}
We solve numerically the above equations by using the shooting method.
In order to utilize this method, one should determine the appropriate asymptote of the solution
at $r \to 0$ for $|V(r)|\approx |V(0)|=\frac{Z e^{2}}{\kappa a}\gg |E|$. This asymptote is
different for orbital numbers $m \ge 0$ and $m < 0$. For $m\geq 0$, by using $f=r^{m+1}\varphi(r)$, $g=r^{m+1}\chi(r)$, and expressing one component in terms of the other, we find the following approximate equation:
\begin{equation}
\varphi''+\frac{2m+1}{r}\varphi'+\left(\frac{V^{2}(0)}{\hbar^{2}v_{F}^{2}}-\frac{(2m+1)}{r^{2}}\right)
\varphi=0,
\end{equation}
whose regular at the origin solution is $\varphi(r)\sim r$. The other component $\chi(r)\sim
{\rm constant}$ for $r\to 0$. Therefore, the radial functions should satisfy the boundary conditions
at the origin
\begin{equation}
\label{initial-condition}
\varphi(0)=0,\ \chi(0)=1.
\end{equation}
For $m<0$, by using $f=r^{-m}\varphi(r)$, $g=r^{-m}\chi(r)$, and proceeding in a similar
way we find that the radial functions $\varphi$ and $\chi$  satisfy the boundary conditions at the
origin which are swapped with respect to those in the case $m \ge 0$,
\begin{equation}
\label{initial-condition2}
\varphi(0)=1,\ \chi(0)=0.
\end{equation}

The numerical integration of Eq.~(\ref{system-eq}) proceeds as follows. We take a ``shot'' from $r=0$
at a fixed value of energy solving the differential equations with the initial conditions (\ref{initial-condition}) or~(\ref{initial-condition2}) and check the behavior of the wave functions
at $r\to\infty$. The latter may tend to $+\infty$ for some values of energy or to $-\infty$ for other values. A physical solution is the solution for which the exponentially growing behavior of the absolute value is absent. We find the corresponding value of the energy of this solution by using the
method of bisections. In our numerical calculations, we use $a=0.05 l_{B}$.

\begin{figure}[hp]
  \centering
  \includegraphics[scale=0.32]{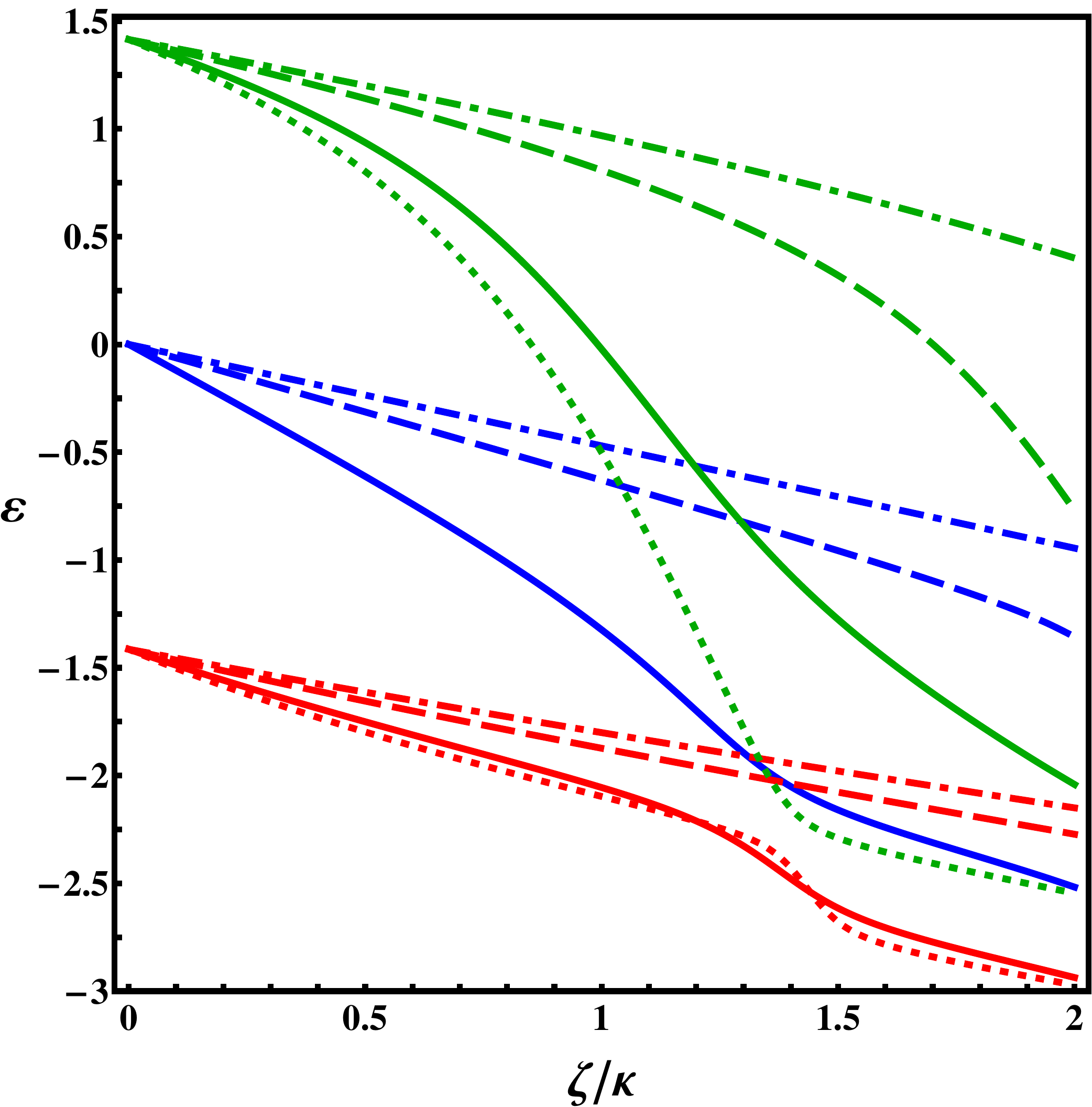}
  \caption{The dimensionless energies $\varepsilon=El_{B}/(\hbar v_{F})$ of electron levels in gapless graphene  in the magnetic field $B=10\,$T as functions of the impurity charge for different orbital numbers $m$. Here the levels are marked $n=+1$, green lines; $n=0$, blue lines; $n=-1$, red lines; $m=-1$, dotted lines (only for $n=+1$ and $n=-1$); $m=0$, solid lines; $m=1$, dashed lines; $m=2$, dot-dashed lines.}
  \label{compare}
\end{figure}

The magnetic field modifies the energy spectrum of electrons in the Coulomb field of the charged
impurity making all continuum states discrete and provides an effective scale given by the magnetic
length. On the other hand, the charged impurity removes the orbital degeneracy of Landau levels transforming the latter into bandlike structures. In Fig.~\ref{compare} we plot in solid lines the dimensionless energies $\varepsilon ={E l_{B}}/{\hbar v_{F}}$ of Landau levels with $m=0$ and different
$n$ as functions of impurity charge in the magnetic field $B=10\,\mathrm T$. The blue solid curve
describes the $m=0$ state of the $n=0$ Landau level, and the red and green solid curves describe the 
$m=0$ level of the lower and upper ``quasicontinua'', respectively,  evolving with the charge of impurity.  As the charge of impurity increases, the blue curve comes close to the red curve. In the absence of  magnetic field, with further increase of the charge of impurity the corresponding bound state would dive into the lower continuum producing a resonance.

According to Fig.~\ref{compare}, the situation is {\it qualitatively} different in
the presence of a magnetic field as the blue curve never crosses the red curve. Instead, typical level repulsions are realized (the well-known avoided crossing theorem \cite{Wigner} forbids a level
crossing for two states with the same symmetry). It is seen that level repulsion occurs only between
the sublevels with the same value of orbital momentum $m$. For example, we clearly see the repulsion
between the levels $n=1,\ m=-1$ and $n=-1,\ m=-1$, as well as the levels $n=0,\ m=0$ and $n=-1,\ m=0$.
States with different quantum numbers $m$ simply cross each other without repulsion. The situation
is similar to that of a quantum electrodynamical system of the finite size \cite{Muller,Greiner}.

\begin{figure}[hpt]
  \centering
  \includegraphics[scale=0.25]{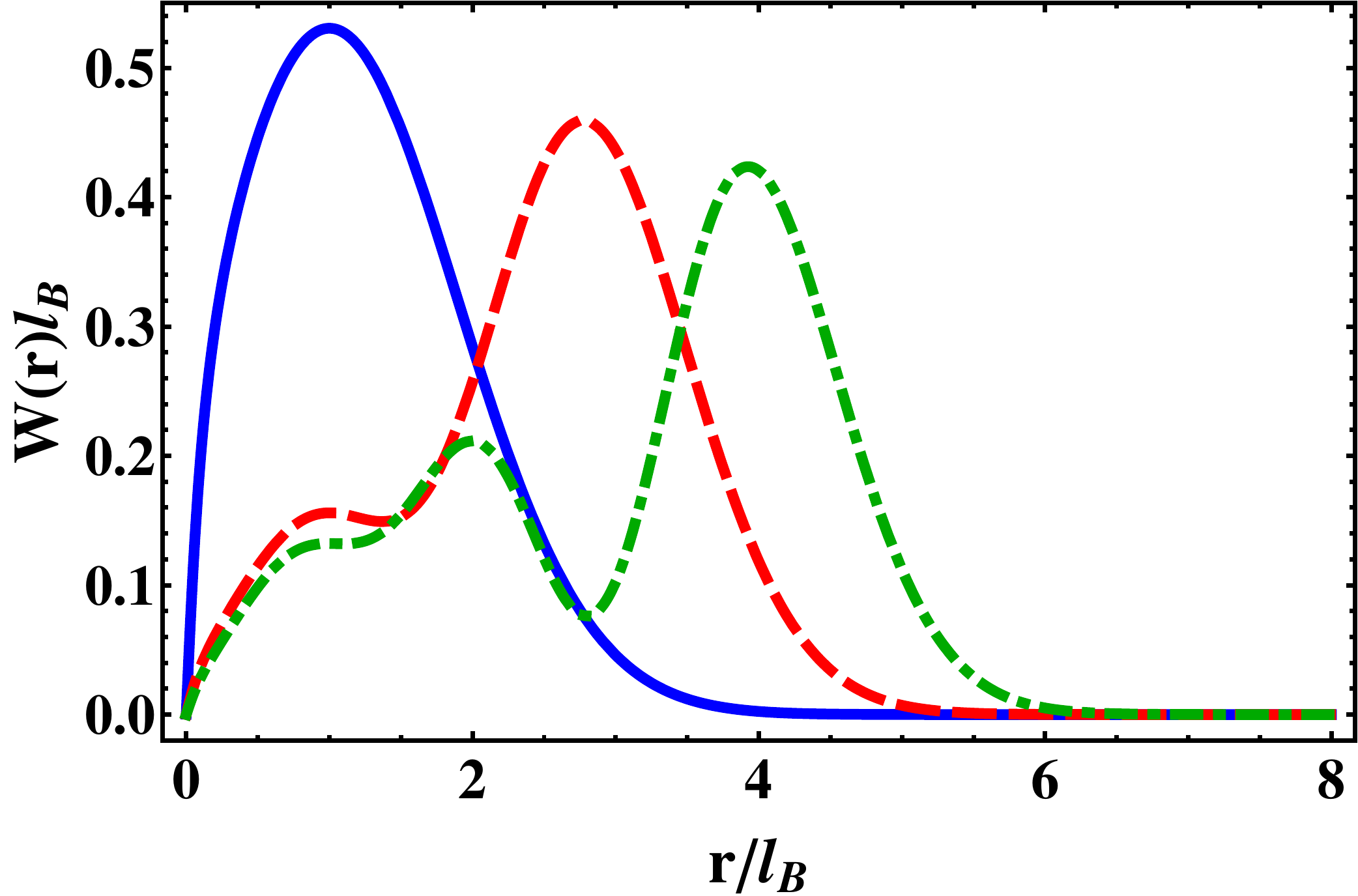}
  \includegraphics[scale=0.25]{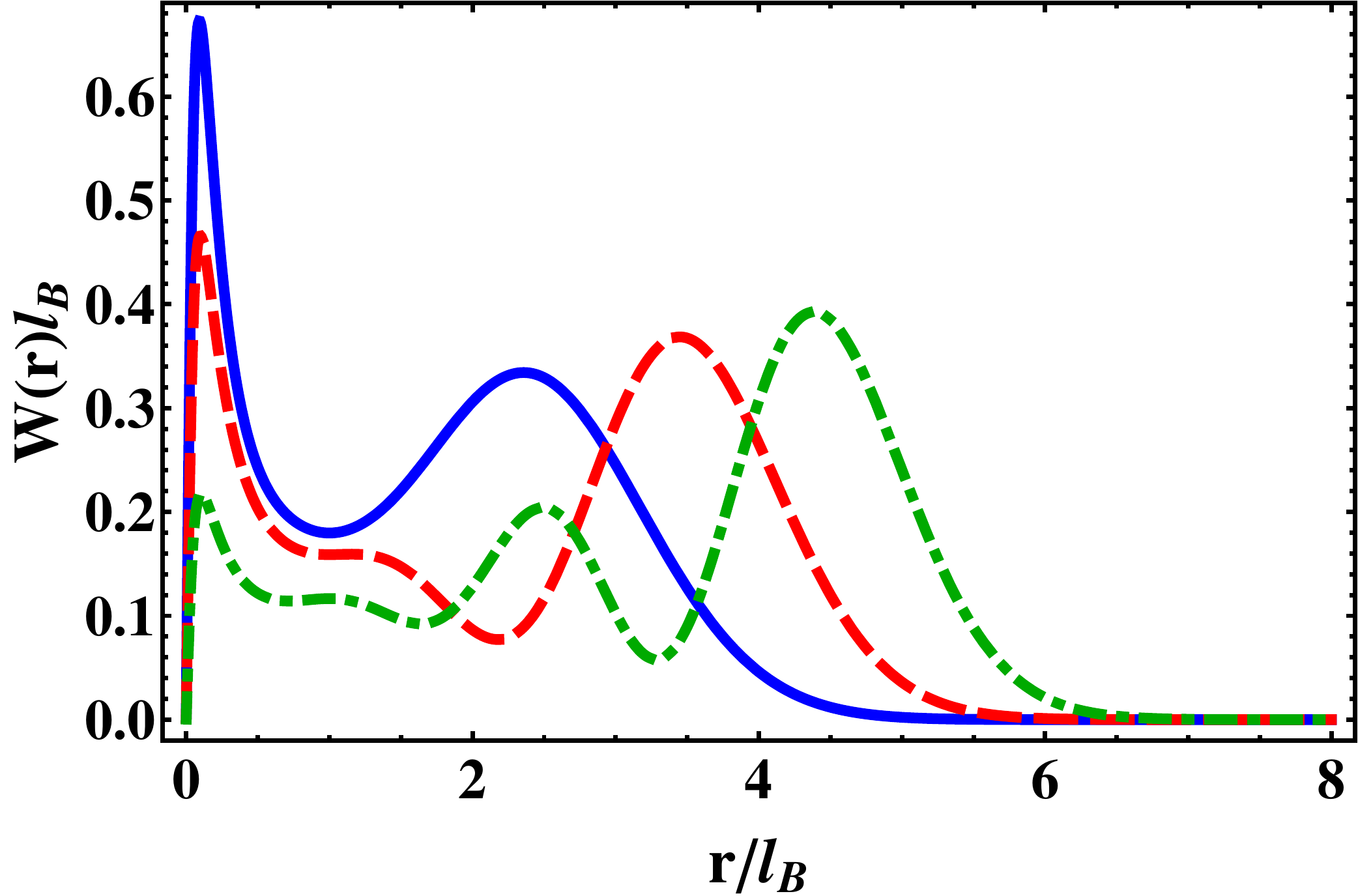}
  \includegraphics[scale=0.25]{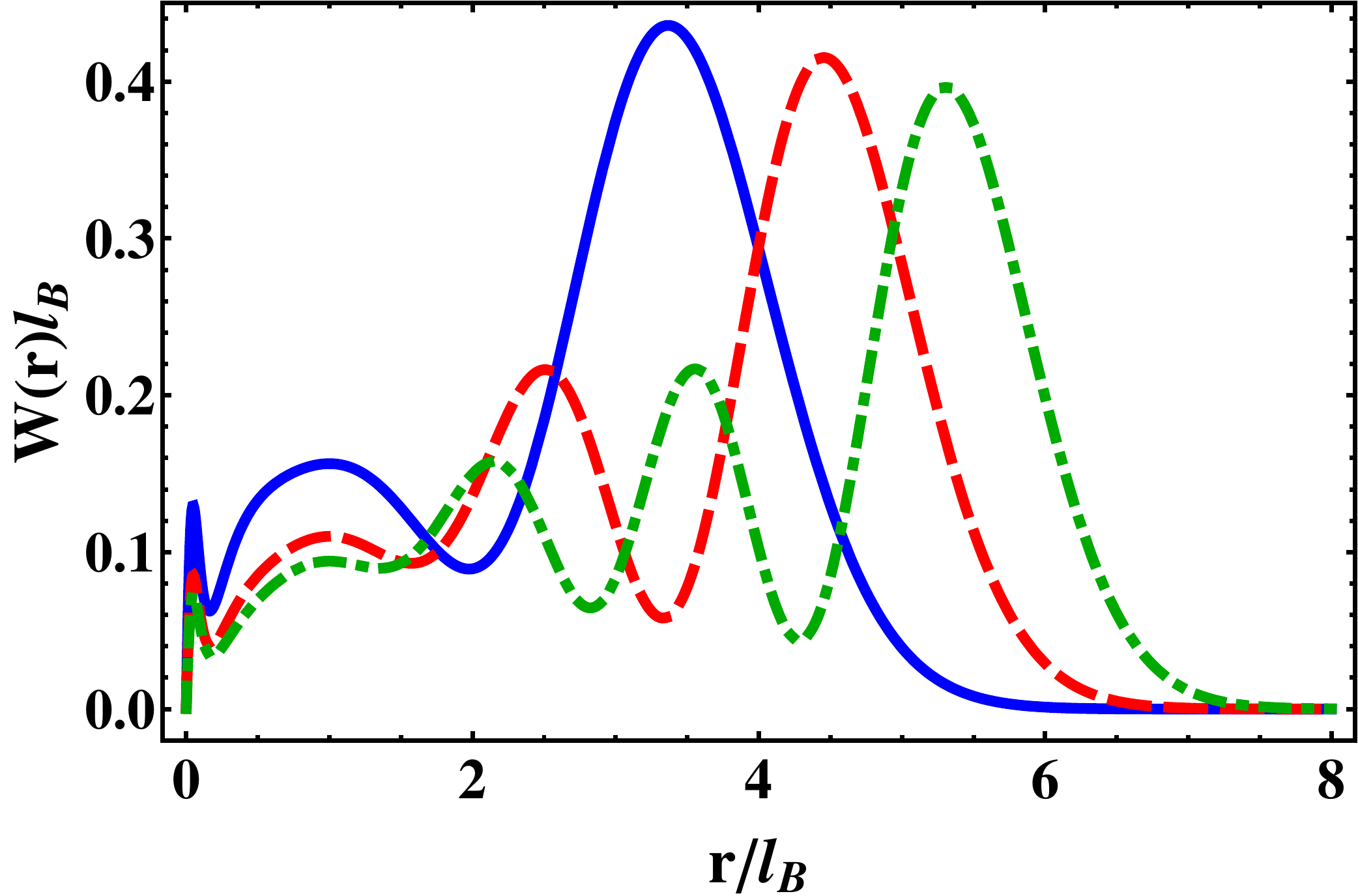}
  \caption{The radial functions of the electron density of the $m=0$ state for the Landau
  levels $n=0$ (blue solid lines), $n=-1$ (red dashed lines), and $n=-2$ (green dot-dashed lines) and
  three different values of the impurity charge: $\zeta/\kappa=0.7$ (left panel), $\zeta/\kappa=1.3$
  (middle panel), $\zeta/\kappa=1.9$ (right panel).}
  \label{WF_mag}
\end{figure}

Figure~\ref{WF_mag} shows the radial distribution function $W(r)=2\pi r|\Psi_{nm}|^2$ for $m=0$ and $n=0,\, -1,\, -2$ states for the three values of the impurity charge $\zeta/\kappa=0.7$, $1.3$, and $1.9$. The second value corresponds to the states in the vicinity of the avoided crossing, see the blue and red solid curves in Fig.~\ref{compare}. For a small charge of the impurity (left panel), the electron density is weakly affected by the impurity and the radial distribution functions of the above mentioned states have one, two, and three maxima, respectively. As the impurity charge increases, all leftmost maxima in $W(r)$ move to the impurity position $r=0$ and attain their maximal values at $\zeta/\kappa\approx 1.3$ (middle panel). In addition, a new maximum appears on the blue solid curve (as well as additional maxima on the other two curves), and the radial distribution function of the $n=0$ level begins to look qualitatively like the radial distribution function of the $n=-1$ level with two maxima.

Further, the middle panel implies that the peak in the radial distribution function of the $n=0$ level
near the impurity is redistributed among the $m=0$ states of the $n=-1,-2, ...$ Landau levels. Obviously, this is an analog of the phenomenon of the diving into continuum for a supercritical charge in the
absence of a magnetic field. In the latter case, the lowest bound state dives into the lower continuum producing
a resonance whose  wave function can be considered as redistributed over the lower continuum states with energies of the order of the resonance width $\gamma$.  All wave functions from this region have an additional sharp peak near the origin. As we see, when magnetic field is present, there is a similar redistribution of the profiles of radial distribution functions near the impurity (note that
as the impurity charge increases, the ``redistribution'' region shifts down to the lower Landau levels). According to the right panel in Fig.~\ref{WF_mag}, the blue curve representing the electron density is now similar to the red dashed curve in the left panel and the red dashed curve is similar to the green dot-dashed curve in the left panel.

So far we did not take into account the screening of a charged impurity due the polarization effects in  graphene to which we turn in the next section.

\section{Tuning the screening of charged impurity with chemical potential}
\label{section-screening}

The strength of the impurity, and consequently the splitting of Landau levels, in the field of charged impurities can be effectively controlled by the gate voltage as was demonstrated in experiment in Ref.~[\onlinecite{states}]. Luican-Mayer \textit{et al}. attribute the variation in the strength of the impurity potential to the screening properties of the 2D electron system. To describe this effect theoretically,
we first consider the polarization function without the impact of the Coulomb impurity. With the 
effects of screening taken into account, the Poisson equation reads
\begin{equation}
\label{Poisson_eq}
\sqrt{-\Delta_{2D}}V^{(0)}_{tot}(\mathbf{x})=-\frac{2\pi Z e^{2}}{\kappa}\delta^{(2)}(\mathbf{x})
-\frac{2\pi e^{2}}{\kappa}\int\!\!d^{2}\mathbf{y}
\Pi^{(0)}(\mathbf{x}-\mathbf{y};\mu)V^{(0)}_{tot}(\mathbf{y}),
\end{equation}
where the polarization function $\Pi^{(0)}(\mathbf{x}-\mathbf{y};\mu)$ is calculated by using the
wave functions in the absence of impurity. Notice the presence of the pseudodifferential operator $\sqrt{-\Delta_{2D}}$ in the equation above, which is necessary  to correctly describe
the Coulomb interaction in a dimensionally reduced electrodynamic system \cite{reduced}.

\begin{figure}[ht]
  \centering
  \includegraphics[scale=0.77]{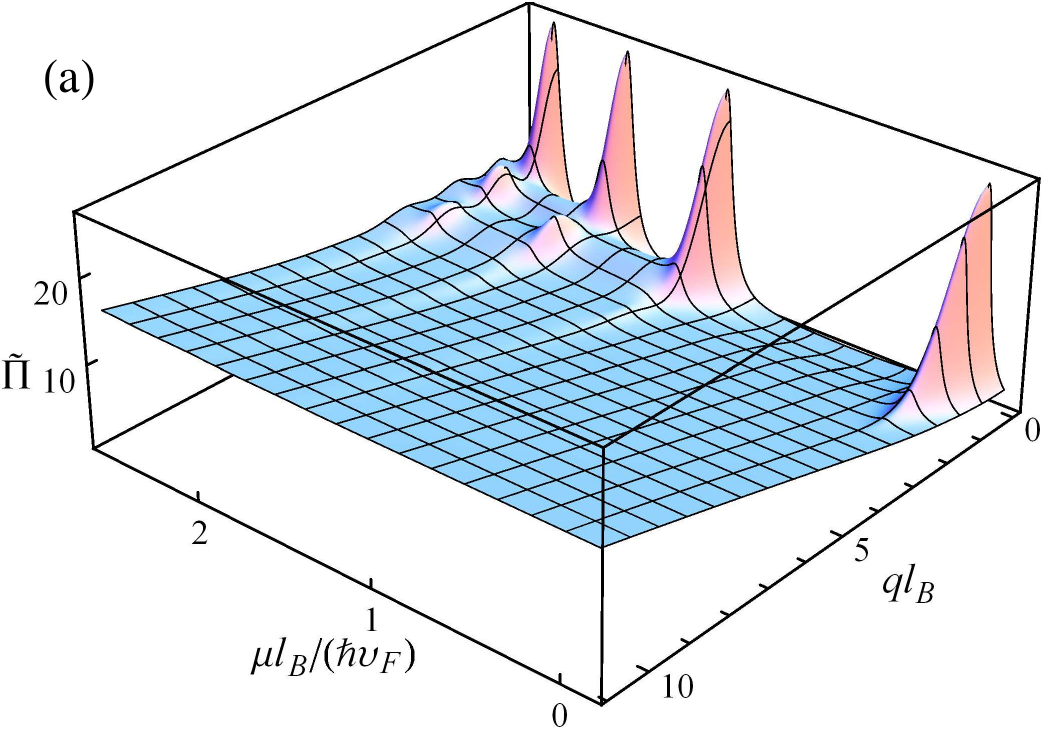}\qquad
	\includegraphics[scale=0.65]{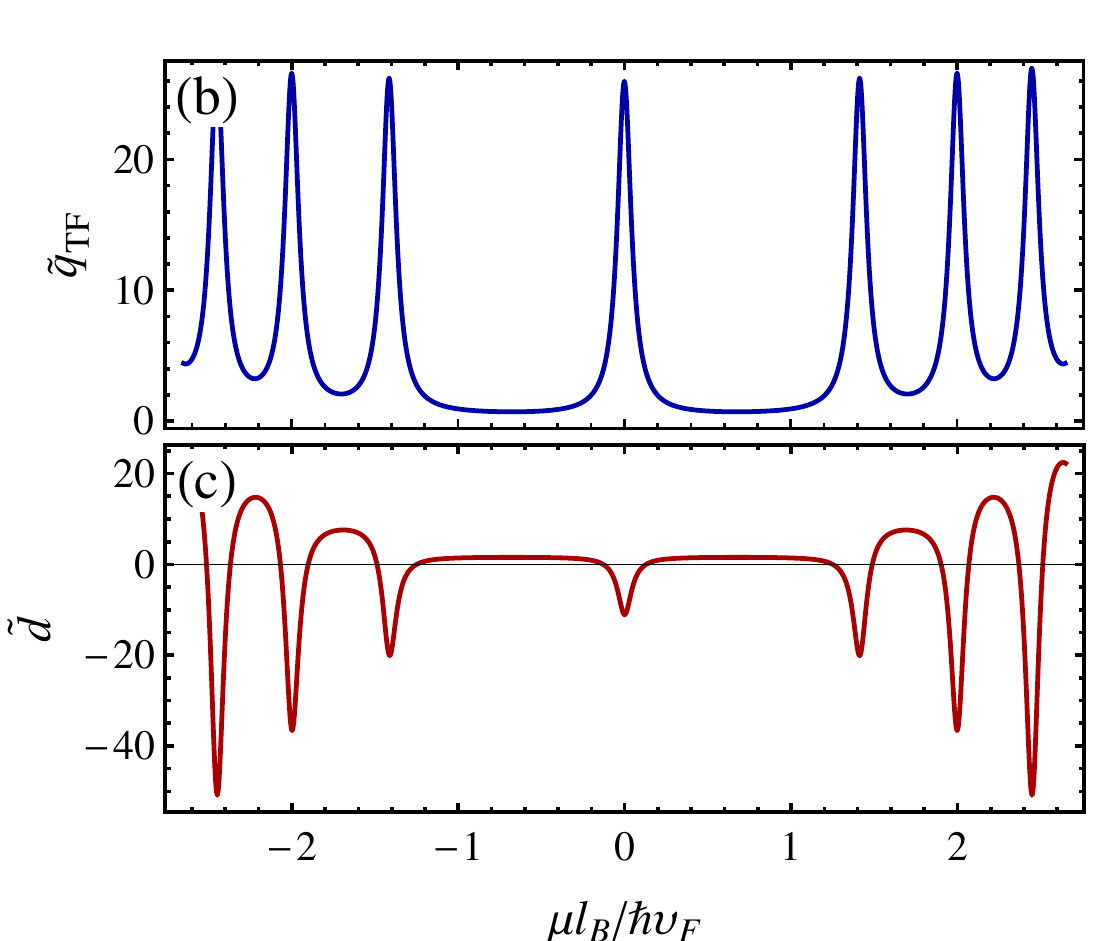}
  \caption{Dimensionless polarization function  $\tilde\Pi=(4\pi\hbar v_{F}l_{B}/N_{f})\Pi(0,q;\mu)$
  as a function of the chemical potential and the wave vector (left panel) and the two coefficients $\tilde q_{\rm TF}=(2\kappa\hbar v_Fl_B/e^2N_f)q_{\rm TF}$, $\tilde d=(2\kappa\hbar v_F/e^2N_fl_B)d$ of its expansion~(\ref{pol_long_wavelength}) at small wave vectors (right panel).}
  \label{pol-function}
\end{figure}

\begin{figure}[ht]
  \centering
  \includegraphics[scale=0.42]{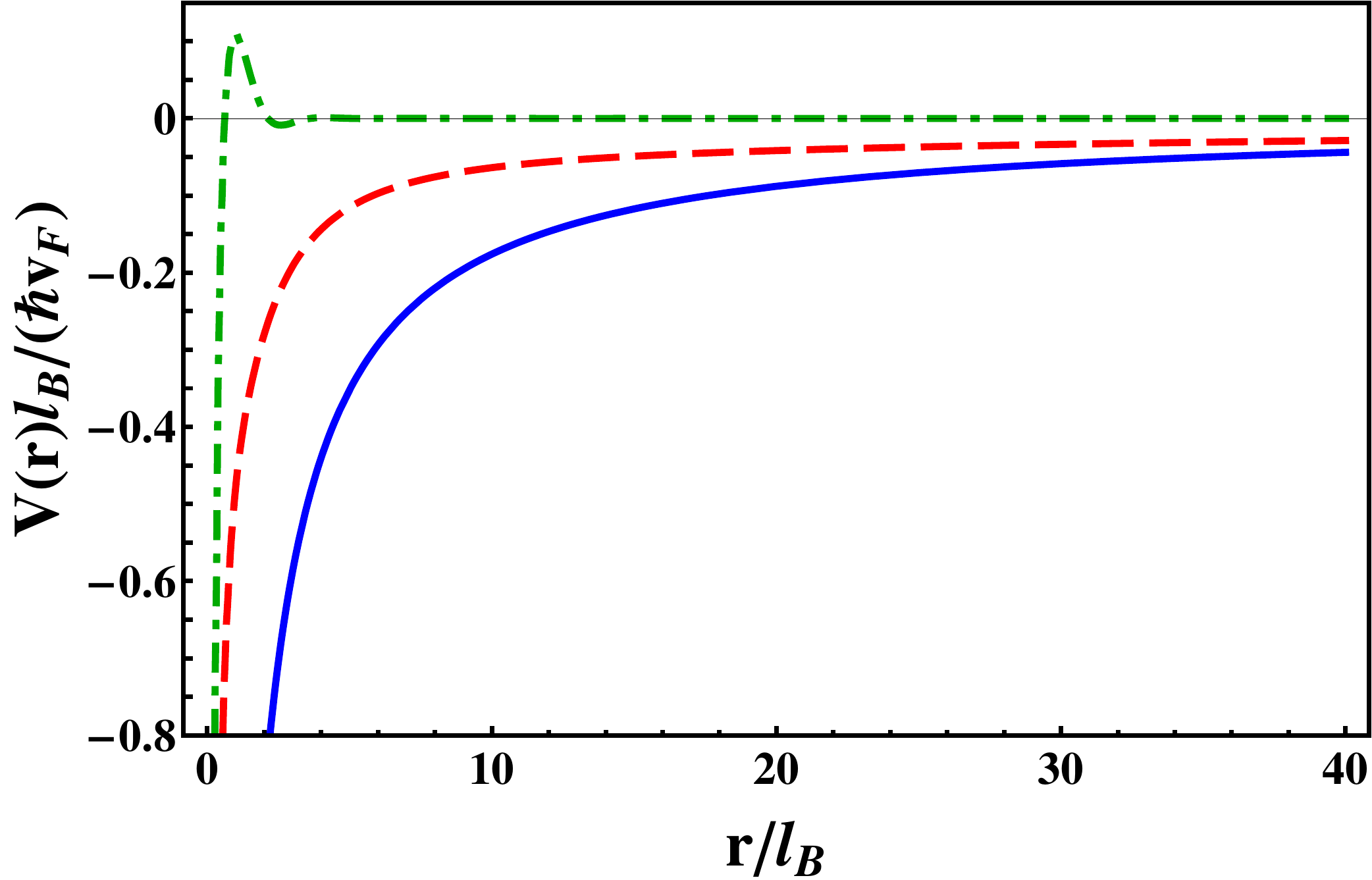}
  \caption{The unscreened regularized Coulomb potential (blue solid line) and the screened potential of the impurity as a function of $r/l_{B}$ in the cases where the chemical potential is situated between Landau levels (red dashed line) and lies inside the zeroth Landau level (green dash-dotted line).}
  \label{pot_scr}
\end{figure}

In order to show that the effective equation for planar charge density distribution has form (\ref{Poisson_eq}), let us start with the  Coulomb potential in momentum space in three dimensions $\tilde{V}(\mathbf{q},q_{z})\sim \frac{1}{\mathbf{q}^{2}+q_{z}^{2}}$, where $\mathbf{q}$ is the
planar momentum and $q_z$ the third component of momentum. To find the effective potential for a
planar distribution of charges  we should integrate over $q_{z}$. Then we find that the effective
potential equals $\tilde{V}_{eff}(q)\sim \frac{1}{|\mathbf{q}|}$. Obviously, the corresponding
potential in  coordinate space satisfies a two-dimensional equation (\ref{Poisson_eq}) with zero polarization function. Since Eq.(\ref{Poisson_eq}) is algebraic in momentum space,
\begin{equation}
\label{alg_eq}
\left(q+\frac{2\pi e^{2}}{\kappa}\Pi^{(0)}(0,q;\mu)\right)V^{(0)}_{tot}(q)=-\frac{2\pi Z e^{2}}{\kappa},
\end{equation}
the potential in coordinate space is easy to find:
\begin{equation}
\label{potential_screened}
V^{(0)}_{tot}(\mathbf{x})= -\frac{Z e^{2}}{\kappa}\int\frac{d^2q}{2\pi}\frac{\exp(i\mathbf{q}\mathbf{r})}
{|\mathbf{q}|+\frac{2\pi e^{2}}{\kappa}\Pi^{(0)}(0,q;\mu)}=
 -\frac{Z e^{2}}{\kappa}\int\limits_{0}^{+\infty}\!\!dq
\frac{q\ J_{0}(q|\mathbf{x}|)}{q+\frac{2\pi e^{2}}{\kappa}\Pi^{(0)}(0,q;\mu)}.
\end{equation}
The static polarization function at zero temperature has the form \cite{Gusynin}
\begin{equation}
\label{polarization}
\Pi^{(0)}(0,q;\mu)=\frac{N_{f}}{4\pi l_{B}^{2}}\left\{\sum\limits_{n=0}^{n_{c}}\sum\limits_{\lambda=\pm}Q^{\lambda\lambda}_{nn}
\left(\frac{q^2l^2_B}2\right)\delta_{\Gamma}(\mu-\lambda M_{n})-\underset{\lambda n\neq\lambda' n'}{\sum\limits_{n,n'=0}^{n_{c}}\sum
\limits_{\lambda,\lambda'=\pm}}Q^{\lambda\lambda'}_{nn'}
\left(\frac{q^2l^2_B}2\right)\frac{\theta_{\Gamma}(\mu-\lambda M_{n})
-\theta_{\Gamma}(\mu-\lambda' M_{n'})}{\lambda M_{n}-\lambda' M_{n'}}\right\},
\end{equation}
where $M_{n}=\frac{\hbar v_{F}}{l_{B}}\sqrt{2n}$ are the Landau level energies, and we
introduced the ultraviolet cutoff $n_c$ because of the divergence of the sum over the Landau levels,
which is estimated to be $n_c=10^4/B[T ]$ due to finiteness of the bandwidth.
As in experiment \cite{states}, we consider the system of two superposed graphene layers twisted
away from Bernal stacking by a large angle. This does not affect the spectrum of single-layer graphene
but results in an additional twofold layer degeneracy: the factor $N_{f}=2_{s}2_{l}=4$ takes into account spin degeneracy and the presence of a second graphene layer.  In experiment, this  setup ensures  reducing
the random potential fluctuations due to substrate imperfections. The smeared delta function $\delta_{\Gamma}(x)=\frac{\Gamma}{\pi}\frac{1}{x^{2}+\Gamma^{2}}$ and the step function
$\theta_{\Gamma}(x)=\frac{1}{2}+\frac{1}{\pi}\arctan\left(\frac{x}{\Gamma}\right)$ account for the finite width of Landau levels, and the functions $Q^{\lambda\lambda'}_{nn'}(y)$ are defined as
\begin{equation}
Q^{\lambda\lambda'}_{nn'}(y)=e^{-y}y^{|n-n'|}\left(\sqrt{\frac{(1+\lambda\lambda'\delta_{0,n_{>}})
n_{<}!}{n_{>}!}}L^{|n-n'|}_{n_{<}}(y)+\lambda\lambda'(1-\delta_{0,n_{<}})\sqrt{\frac{(n_{<}-1)!}
{(n_{>}-1)!}}L^{|n-n'|}_{n_{<}-1}(y)\right)^{2},
\label{Q-function}
\end{equation}
where $n_<=\min(n,n')$, $n_>=\max(n,n')$ and $L_n^m(y)$ are the generalized Laguerre polynomials.
The first term in Eq.~(\ref{polarization}) describes the contribution from the intralevel transitions
while  the second term represents contributions from the interlevel transitions. For small width of Landau
levels the first term looks like a sequence of delta functions  and  contributes only when the chemical potential lies inside Landau levels.
At small wave vectors ($q\ll l_B^{-1}$) the polarization function~(\ref{polarization}) behaves as \cite{Gusynin}
\begin{equation}
\Pi^{(0)}(0,q;\mu)\simeq\frac{\kappa}{2\pi e^2}(q_\text{TF}+d q^2),
\label{pol_long_wavelength}
\end{equation}
where
\begin{equation}
q_\text{TF}=\frac{e^2N_f}{\kappa l^2}\sum_{n=0}^{n_c}\sum_{\lambda=\pm}(2-\delta_{0n})\delta_\Gamma(\mu-\lambda M_n)
\label{q_TF}
\end{equation}
is the Thomas-Fermi wave vector which determines the strength of the long-wavelength screening,
and parameter $d$ is given by
\begin{equation}
d=-\frac{e^2N_f}{2\kappa}\sum_{n=0}^{n_c}\sum_{\lambda=\pm}(4n+\delta_{0n})
\delta_\Gamma(\mu-\lambda M_n)-\frac{e^2N_fl}{2\sqrt2\kappa\hbar v_F}\sum_{n=0}^{n_c-1}
\sum_{\lambda,\lambda'=\pm}\frac{\theta_\Gamma(\mu-\lambda M_{n+1})-\theta_\Gamma(\mu-\lambda' M_n)}
{(\lambda\sqrt{n+1}-\lambda'\sqrt n)^3}.
\label{a_coeff}
\end{equation}
Figure~\ref{pol-function} illustrates the dependence of the static polarization function~(\ref{polarization}) and its two leading long-wavelength terms~(\ref{q_TF}) and~(\ref{a_coeff}) on the chemical potential.
We plot for comparison the unscreened potential and the screened potential (\ref{potential_screened}) of
the impurity in Fig.~\ref{pot_scr}. Let us consider the case where the chemical potential is situated between Landau levels. Then the Thomas-Fermi wave vector~(\ref{q_TF}) is close to zero [Figs.~\ref{pol-function}(a) and~\ref{pol-function}(b)] and the Coulomb potential of the impurity is
weakly screened, although even in this case graphene contributes to the total dielectric function at large and intermediate momenta, which effectively diminishes the charge of the impurity and the screened potential. Indeed, while the screened potential tends to its bare value at $r\to\infty$, it is weakened for small and intermediate distances (see the red dashed line in Fig.~\ref{pot_scr}). On the other hand, when the chemical potential lies inside any given Landau level, the screening works much more effectively due to large $q_\text{TF}$ (see the green dash-dotted line in Fig.~\ref{pot_scr}) providing an excellent means of controlling the effective charge of impurity by the gate voltage. Moreover, the coefficient $d$ in Eq.~(\ref{a_coeff}) in this case is negative [see Fig.~\ref{pol-function}(c)] which means that $\Pi^{(0)}(0,q;\mu)$ has a nonmonotonic momentum dependence with a peak at $q=0$. This behavior of the polarization function leads to the oscillations of the screened potential (green dash-dotted line in Fig.~\ref{pot_scr}) with the sign change (i.e., the overscreening of the Coulomb potential) at intermediate distances of the order of several magnetic lengths.

\begin{figure}[hpt]
  \centering
  \includegraphics[scale=0.42]{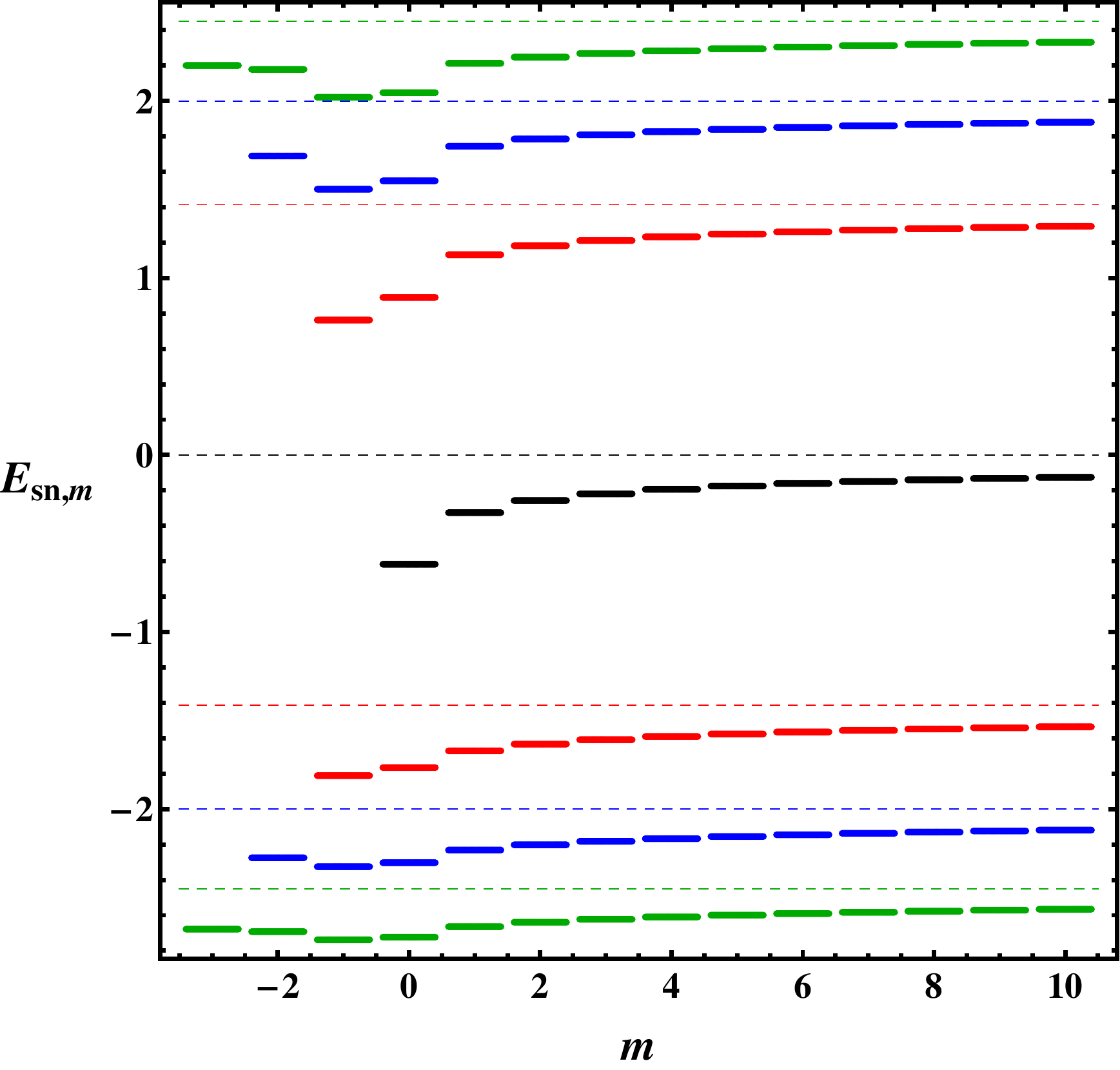}
  \caption{The energies of Landau levels  $n=0,\pm 1, \pm 2, \pm 3$ split with $m$ in the field of screened impurity in a magnetic field, plotted in units of ${\hbar v_{F}}/{l_{B}}=81.2$\,meV in the case where the chemical potential lies between Landau levels. The bare charge of the impurity is $Z=+2$, the dielectric constant of the substrate is $\kappa=2.5$, and the magnetic field is $B=10$\,T.}
  \label{lev_mag}
\end{figure}

Let us now consider the electron states. By numerically solving the Dirac equation with the screened potential~(\ref{potential_screened}) using the same procedure as in the previous section, we find the electron spectrum which shows that the Landau levels are shifted and split into sublevels with different values of $m$. We perform calculations for the doubly charged impurity $Z=+2$ and the dielectric constant
due to silicon substrate $\kappa=(1+\kappa_{sub})/{2}=2.5$. The energies of electron sublevels for several first Landau levels are displayed in Fig.~\ref{lev_mag}.

In general, the splitting of Landau levels with $m$ can be determined only numerically. However, for
large $m$, it can be found in perturbation theory. Unperturbed wave functions of the $n=0$ Landau
level of the electron in a magnetic field are given by
\begin{equation}
\Psi^{(0)}_{0m}(\mathbf{r})=\frac{e^{-\frac{r^{2}}{4l_{B}^{2}}}}{l_{B}\sqrt{2\pi m!}}\left[
\begin{array}{c}
0\\
\left(\frac{r^{2}}{2l_{B}^{2}}\right)^{m/2}e^{-i m \theta}
\end{array}
\right].
\end{equation}
The lowest Landau level splitting can be estimated as follows:
\begin{equation}
\delta E_{0m}=\langle 0m|V_{tot}(r)|0m\rangle=\int d^{2}\mathbf{r}\ \Psi^{(0)\dagger}_{0m}V_{tot}(r)\Psi^{(0)}_{0m}=\int\limits_{0}^{\infty}
W(r)V_{tot}(r)dr,
\label{LLL-splitting}
\end{equation}
where $W(r)=2\pi r |\Psi^{(0)}_{0m}(\mathbf{r})|^{2}=\frac{1}{2^{m}m! l_{B}^{2m+2}}e^{-r^{2}/(2l_{B}^{2})}r^{2m+1}$ is the radial distribution
function. For large $m$,
it has a high and narrow peak at the distance $r_{m}^{peak}=l_{B}\sqrt{2m+1}$ determined by solving
the equation $W'(r)=0$. For large orbital momenta $m$, the level splitting could be
estimated as $V_{tot}(r_{m}^{peak})$, which approximates  the exact value of
$\delta E_{0m}$ with one percent error for states with $m \ge 10$.

\begin{figure}[hpt]
  \centering
  \includegraphics[scale=0.33]{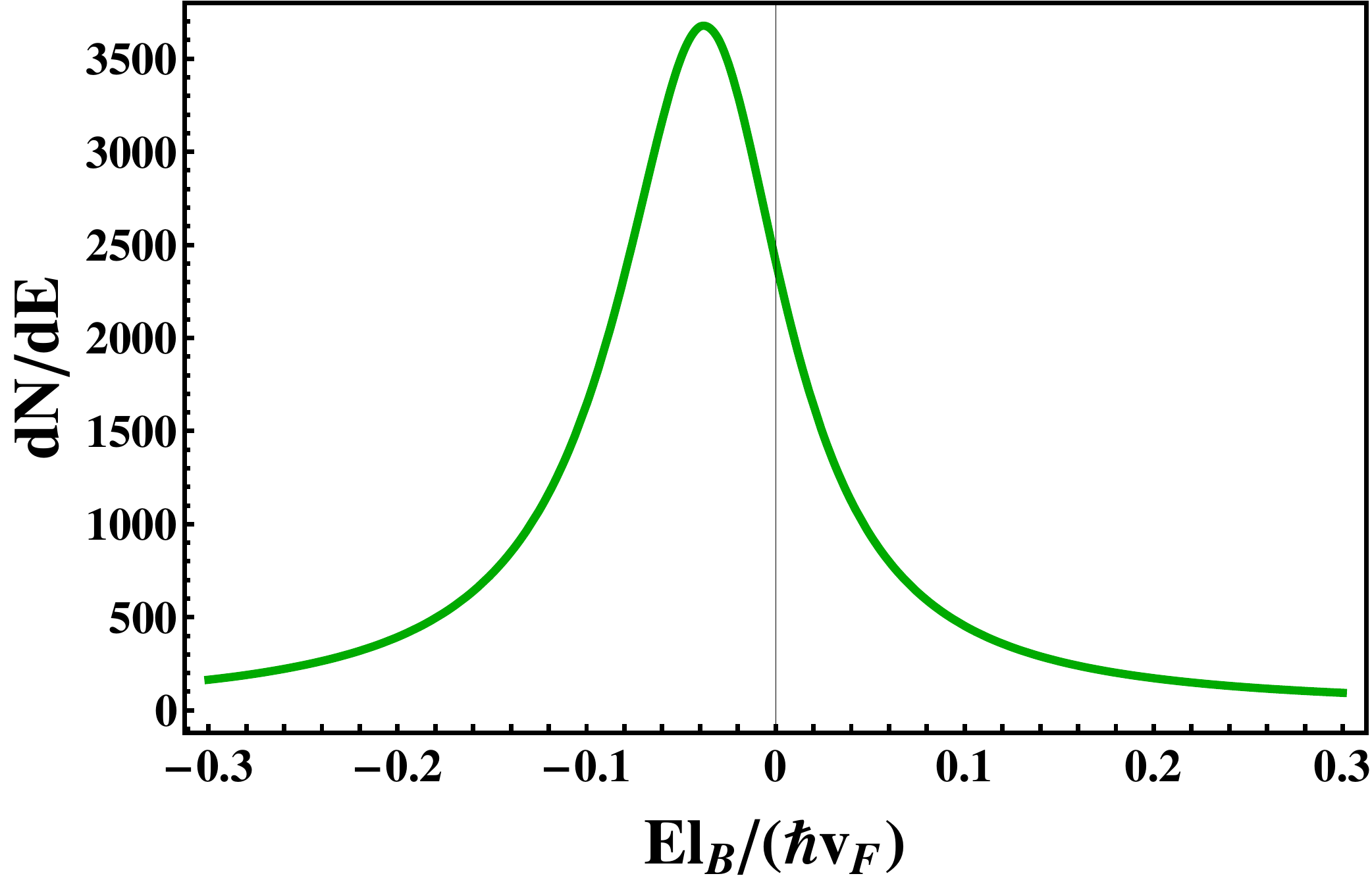}
  \includegraphics[scale=0.33]{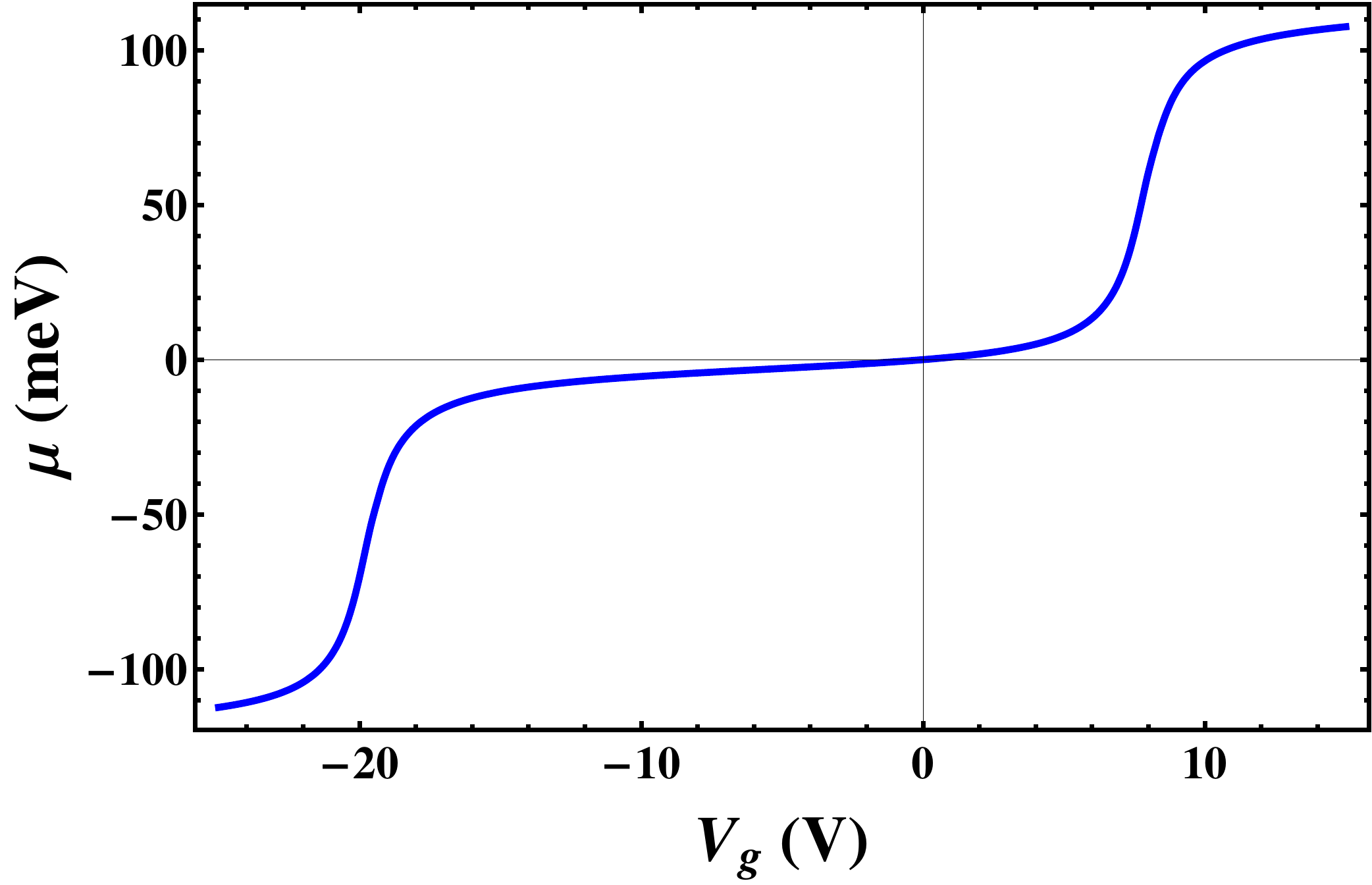}
  \caption{The integral density of states on the zeroth Landau level in graphene with the charged impurity $Z=+2$ (left panel). The dependence of chemical potential on the applied gate voltage (right panel).}
  \label{peak}
\end{figure}

Due to the additional degeneracy $g_{s}g_{v}g_{l}=8$ related to the spin, valley,
and layer degrees of freedom ($g_{s}=g_{v}=g_{l}=2$), the complete degeneracy of a Landau level per
unit area equals $n_{0}(B)=g_{s}g_{v}g_{l}(Be/hc)\approx 2\cdot 10^{12}\ {\rm cm}^{-2}$ in
the magnetic field $B=10$\,T. Because of the finiteness of the sample sizes used in experiment \cite{states}, we also consider the graphene sheet of a finite area taken $500\times 500\,$nm in our calculations. Therefore, there are $N=\frac{n_{0}(B)S}{g_{s}g_{v}g_{l}}\approx 600$ sublevels with different orbital numbers $m$ on each Landau level, so that the maximum orbital number for $n$-th Landau level reads $m_{max}=N-|n|-1$. Broadening these sublevels with the width $\Gamma=0.05{\hbar v_{F}}/{l_{B}}$ used
in Ref.\,[\onlinecite{states}], we plot the integral density of states for the zeroth Landau level
\begin{equation}
\label{sum_density}
\frac{dN(E)}{dE}=\sum\limits_{m=0}^{m_{max}}\delta_{\Gamma}(E-\delta E_{0m})
\end{equation}
in the left panel of Fig.~\ref{peak}. Since the charged impurity affects most strongly the states of
Landau levels with small values of $m$, we use the energies of the exact solutions of the Dirac equation
for the states with $m \le 6$ in Eq.~(\ref{sum_density}) and energies of the rest of the states are computed according to Eq.~(\ref{LLL-splitting}), whose inaccuracy is smaller than one percent for these states.
This panel shows that the zero Landau level is slightly displaced below the origin; thus a large part of this level is already filled at zero gate voltage, i.e., at zero chemical potential.
Therefore, the dependencies $\mu(V_{g})$ and $\Pi^{(0)}(0,q;\mu(V_{g}))$ have a significant asymmetry with respect to gate voltage.

According to Refs.~[\onlinecite{Novoselov, Novoselov2005}], the typical thickness of the SiO$_{2}$
substrate layer is $t\approx 300\,\rm nm$. Then, the carrier density is related to the gate voltage as
$n=\kappa V_{g}/(te)\approx 7\times 10^{10} V_{g}[{\rm V}]\ {\rm cm}^{-2}$, where $e$ is the electron charge and $\kappa$ is dielectric constant due to the SiO$_{2}$ substrate. Taking into account the size of graphene
sheet, the number of electrons appearing in graphene due to the applied gate voltage equals $\Delta N=nS\approx 175\cdot V_{g}[V]$. On the other hand, $\Delta N=g_{s}g_{v}g_{l}\int_{0}^{\mu}\frac{dN}{dE}dE$. Therefore, by integrating expression (\ref{sum_density}) over energy, we obtain the dependence $V_{g}(\mu)=\frac{8}{175}\int_{0}^{\mu}\frac{dN}{dE}dE$. We note that although in general the relationship between the charge density $n$ and the gate voltage $V_g$ is not  strictly linear due to the quantum capacitance effects near the Dirac point, the measurements done in Refs. [\onlinecite{Novoselov,Novoselov2005}] agree with the linear dependence.
We incorporate this relationship in our computations. The inverse dependence is plotted in the right panel
of Fig.~\ref{peak}. This procedure of calculating the dependence of $\mu$ on $V_g$ is easily extended to higher Landau levels.

The local density of states (LDOS) is given by
\begin{equation}
\rho(r, E,\mu)=g_s g_v g_l\sum\limits_{n,m}|\Psi_{nm}(r)|^{2}\delta_{\Gamma}(E-E_{nm}(\mu)),
\end{equation}
where the energies depend on the chemical potential through a screened potential. In experiment [\onlinecite{states}], the electron spectrum was determined by measuring  the differential tunneling conductance $dI/dV_{bias}$ at the tip position as the function of the bias voltage $V_{bias}=(E-E_F)/e$ where the energy $E$ is measured relative to the Fermi level $E_F$ (at zero or sufficiently small temperature $E_F \approx \mu$). Far away from the impurity, the LDOS displays practically unbiased
Landau levels because only wave functions of the states with large $m$ contribute at large distances.
The energies of these states are only weakly shifted so that
$E_{nm}\approx E_{n}^{(0)}={\rm sgn}(n)({\hbar v_{F}}/{l_{B}})\sqrt{2|n|}$ and the wave functions
$\Psi_{nm}(r)\approx \Psi^{(0)}_{nm}(r)$ with good accuracy. In this case one can perform the summation
over the orbital number $m$ and get
\begin{equation}
\rho(r, V_{bias},\mu)=
\frac{g_s g_v g_l}{2\pi l^2_B}\sum\limits_{n=-n_{c}}^{n_{c}}\delta_{\Gamma}(eV_{bias}+\mu-E_{n}^{(0)}).
\end{equation}
Therefore, the LDOS is similar to that when the impurity is absent and the only effect of the latter is
the chemical potential asymmetry due to the shift of the Landau levels. At large distance from the
impurity ($r\gg l_{B}$) the local density of states exhibits the peaks which correspond to unperturbed
Landau levels. They appear at biases $eV_{bias}=E_{n}^{(0)}-\mu$. This is in agreement with the
experimental data (see Fig. 3(a) in Ref. [\onlinecite{states}]).

To account for the reaction of the impurity on the polarization function we proceed
in the following way. We neglect the influence of impurity on the electron wave functions, which keeps
the polarization function translation invariant, but take into account the change of energy levels
due to impurity.

\begin{figure}[hpt]
  \centering
  \includegraphics[scale=0.33]{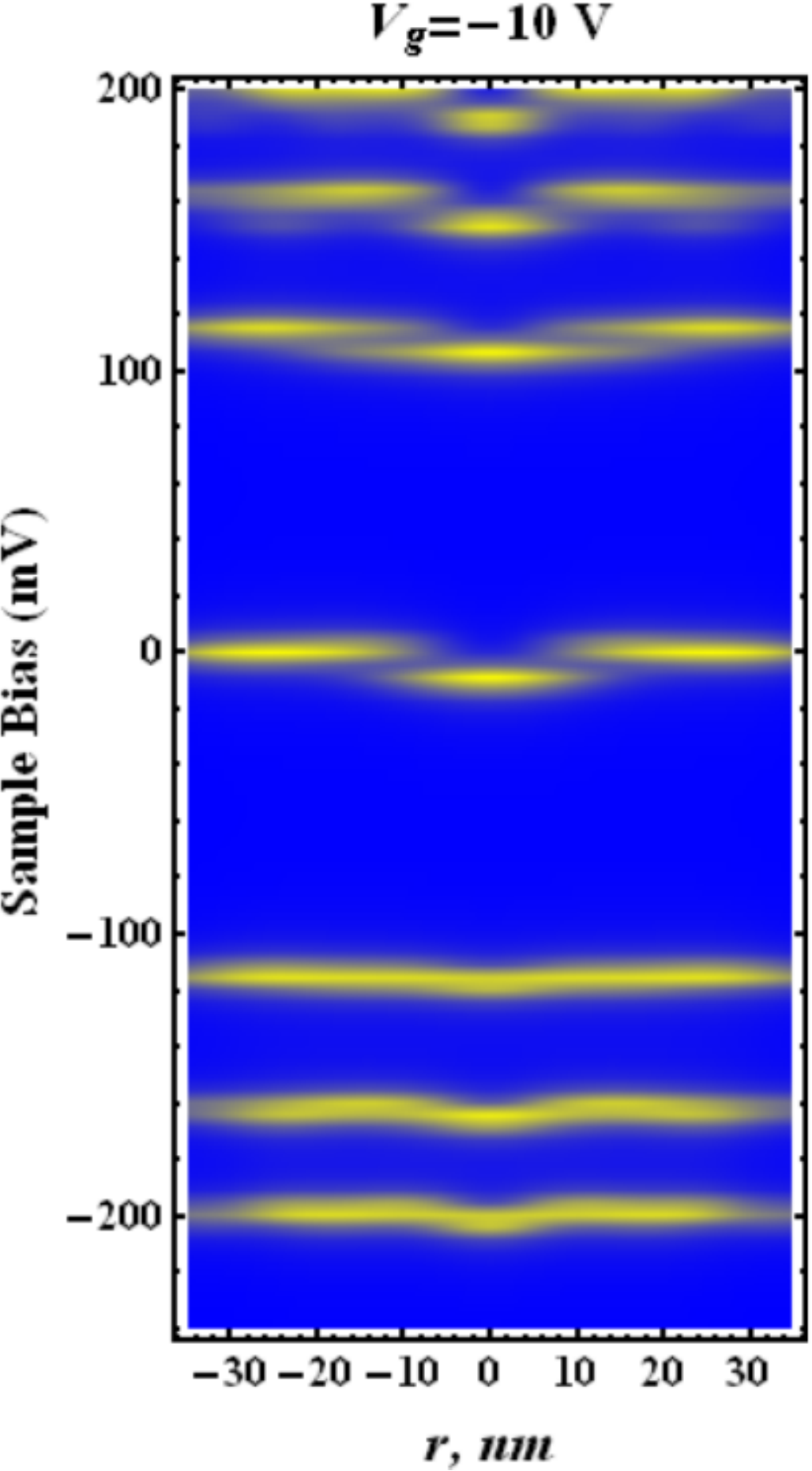}
  \includegraphics[scale=0.33]{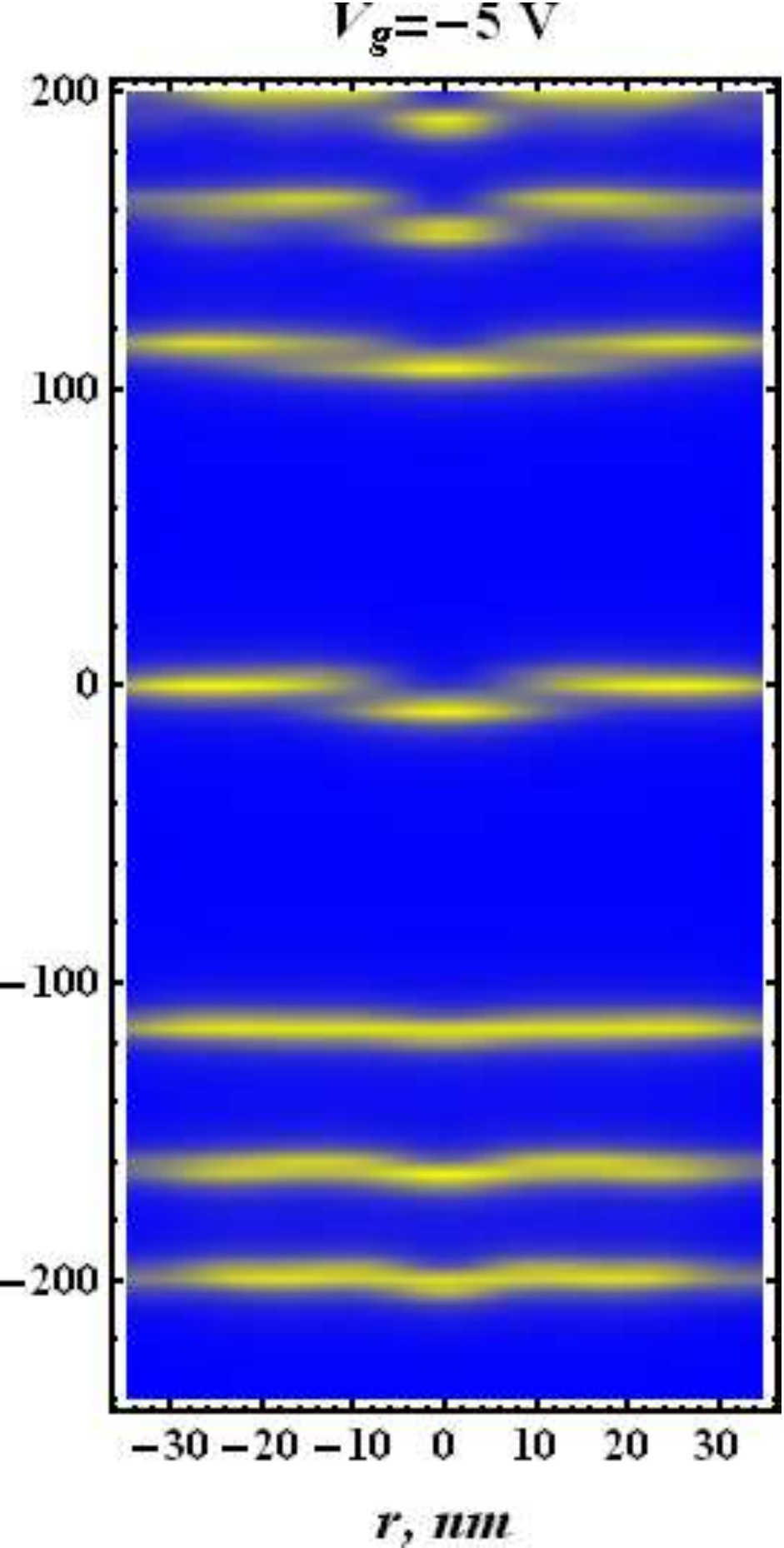}
  \includegraphics[scale=0.33]{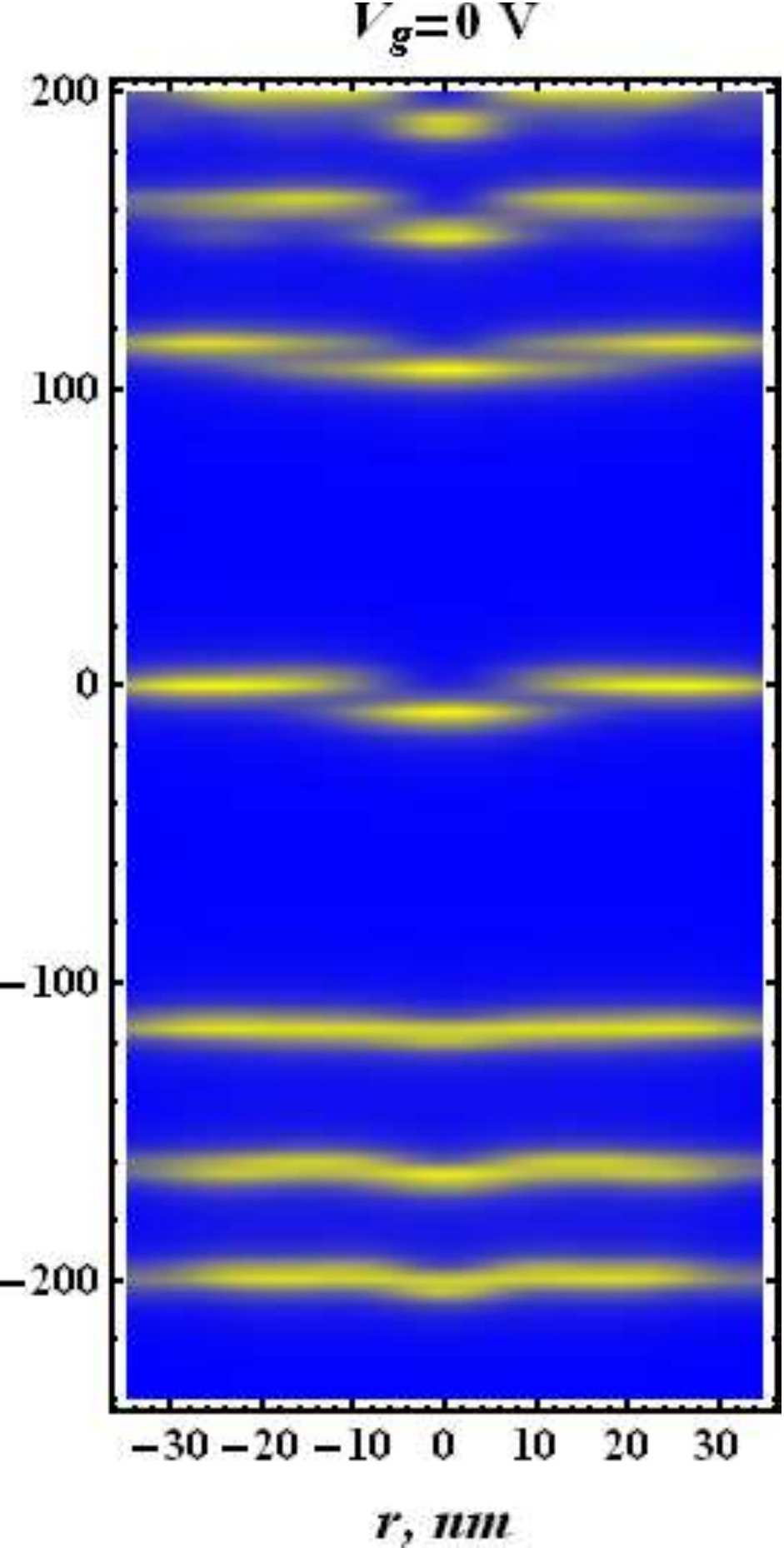}
  \includegraphics[scale=0.33]{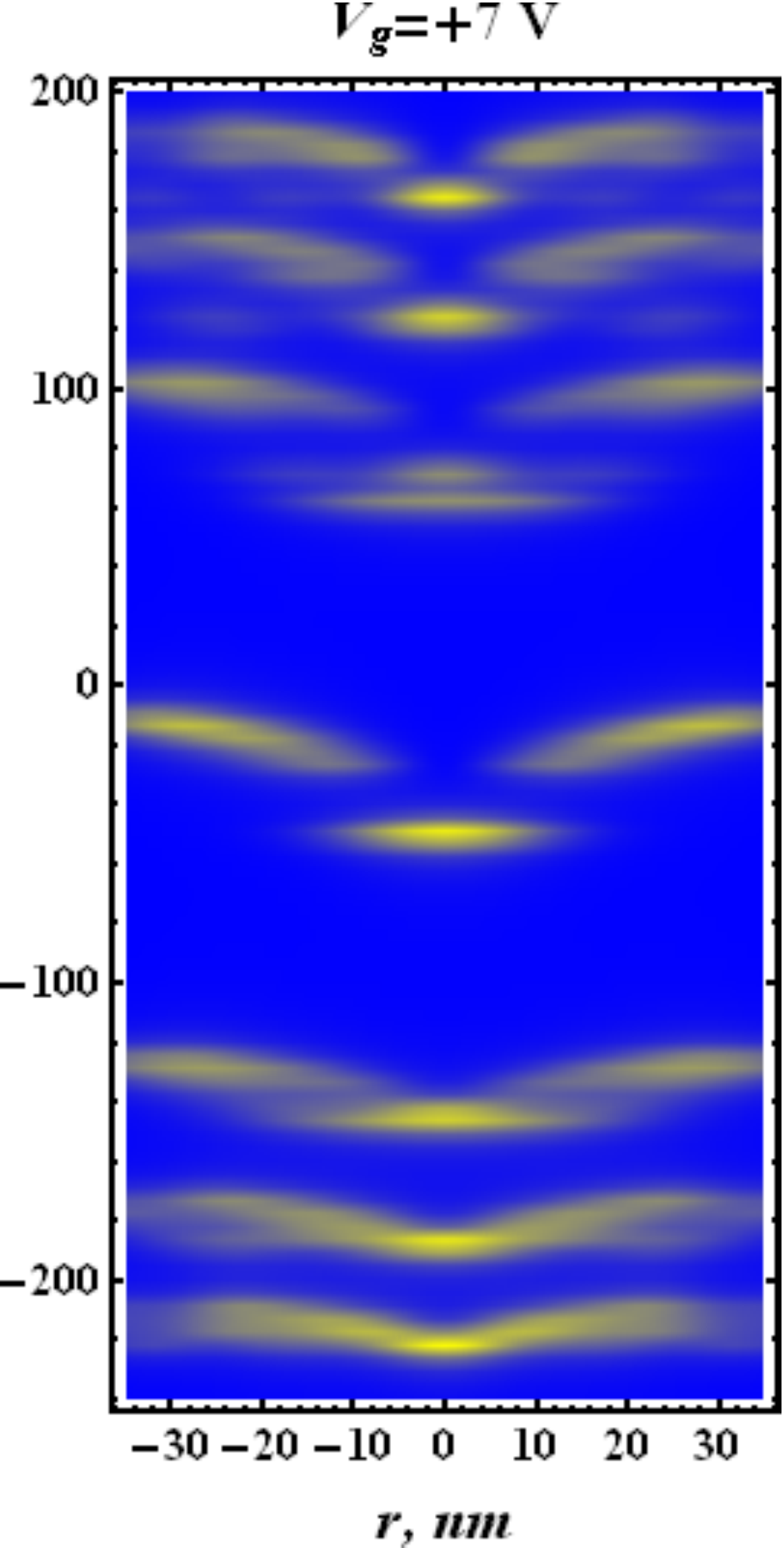}
  \includegraphics[scale=0.33]{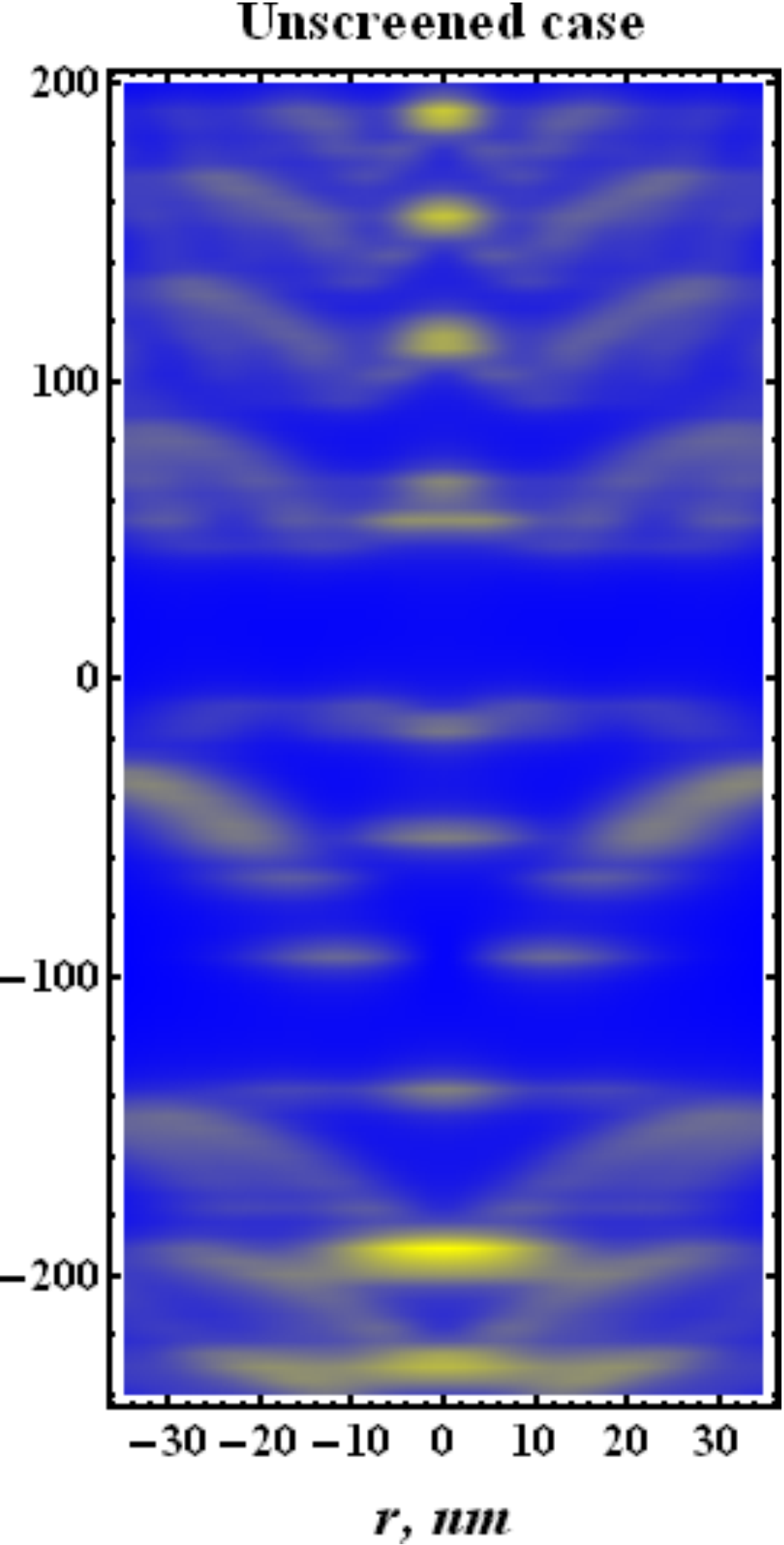}
  \caption{The local density of states is plotted at four values of gate voltage along the line cuts across the impurity. The rightmost panel shows the LDOS in the nonscreened case.
  \label{LDOS}}
\end{figure}

When the impurity is absent, the Landau levels are degenerate in the orbital quantum number and all $g_{s}g_{l}g_{v}N\approx 4800$ electrons
have energies which depend only on the Landau level index $n$. Let us consider the zeroth Landau level. The corresponding contribution from all
its electrons to the polarization function is
$\delta\Pi_{0}=\frac{N_{f}}{4\pi l_{B}^{2}}2Q^{++}_{00}(\frac{q^{2}l^2_B}{2},0)\delta_{\Gamma}(\mu)$. Since the impurity potential splits the
Landau levels into sublevels specified by the orbital
quantum number $m$, each sublevel with fixed $m$ contains only 8 electrons. Naturally, the contribution
of each sublevel must be 600 times smaller than the contribution of the whole level, because the polarization effects are proportional to the
number of electrons which screen the external potential.
The  approximation which, nevertheless, allows us to take into account the shift and broadening of the Landau level due to the impurity
potential is to replace $\delta_{\Gamma}(\mu)\rightarrow\frac{1}{N}\!\!\left.\frac{dN}{dE}\right|_{E=\mu}$, where the derivative $dN/dE$ is
given by Eq.~(\ref{sum_density}), in the first term for $n=0$ in Eq.~(\ref{polarization}):

\begin{eqnarray}
\label{polarization2}
\Pi^{(0)}(0,q;\mu)&=&\frac{N_{f}}{4\pi l_{B}^{2}}\left\{\frac{2Q^{++}_{00}({q^{2}l^2_B}/{2},0)}{m_{max}+1}
\sum\limits_{m=0}^{m_{max}}\delta_{\Gamma}(\mu-\delta E_{0m})+\sum\limits_{n=1}^{n_{c}}\sum\limits_{\lambda=\pm}Q^{\lambda\lambda}_{nn}
\left({q^{2}l^2_B}/{2},0\right)\delta_{\Gamma}(\mu-\lambda M_{n})\right.\nonumber\\
&-&\left.\underset{\lambda n\neq\lambda' n'}{\sum\limits_{n,n'=0}^{n_{c}}\sum\limits_{\lambda,\lambda'=
\pm}}Q^{\lambda\lambda'}_{nn'}\left({q^{2}l^2_B}/{2},0\right)\frac{\theta_{\Gamma}(\mu-\lambda M_{n})
-\theta_{\Gamma}(\mu-\lambda' M_{n'})}{\lambda M_{n}-\lambda' M_{n'}}\right\}.
\end{eqnarray}
In this way, we can incorporate the displacement and broadening of Landau levels due to the
Coulomb field of impurity. Although similar procedures can be performed for several first Landau levels,
the largest contribution is connected with the zeroth Landau level because we consider the case where the chemical potential crosses this
Landau level. This replacement makes the polarization function
asymmetric with respect to the chemical potential, as could be seen from Fig.~\ref{LDOS}.
Using the procedure described above, we compute $V^{(0)}_{tot}(\mathbf{x})$ for four values of gate
voltage. By numerically integrating the Dirac equation with this potential and determining the energies
of several first Landau levels in such a potential, we plot the local density of states in Fig.~\ref{LDOS}.
For comparison with experimental data we take the same values of gate voltage as in
Ref.~[\onlinecite{states}]. It is clearly seen that the impurity is strongly screened when the chemical potential moves within a Landau level (gate voltages $V_g=-10$\,V, $-5$\,V, $0$\,V) with progressive level filling when $V_g$ increases, and screening is strongly diminished when the chemical potential is situated between Landau levels ($V_g=+7$\,V). We also plot the LDOS when the polarization effects are switched off
(the rightmost panel). It shows that the level splitting is much more significant comparing to the case where the polarization effects are taken into account.

\section{Summary}
\label{section-conclusion}

Motivated by a recent experimental study of the Dirac equation for quasiparticles in the field of a
charged impurity in graphene in a magnetic field, we calculated the corresponding electron states
in the continuum model and constructed the local density of states. In the presence of a charged impurity, degenerate Landau levels convert into bandlike structures due to lifting the orbital degeneracy.
For zero chemical potential, as the charge of impurity increases, the energy level with quantum numbers $n=0$, $m=0$ comes close to the highest energy of the level $n=-1$. In the absence of magnetic field, the corresponding bound state would dive into the lower continuum with further increase of the charge of impurity producing a resonance. The situation is qualitatively different in the presence of a magnetic
field as the energy curves with the same orbital momenta $m$ never cross. Our calculations clearly demonstrate this phenomenon of the level repulsion  between the sublevels with the same  $m$ and the formation of a quasiresonance state when the impurity charge exceeds a critical value. In such a case
we observe a redistribution of profiles of radial distribution functions with the same orbital
momentum among lower Landau levels $n\le-1$.

Experimentally, it was shown that the strength of a charged impurity and splitting of Landau sublevels
with different orbital momenta in a magnetic field can be very effectively tuned by a gate voltage.
To describe this phenomenon theoretically it is crucial to take into account the polarization
in a magnetic field in the presence of chemical potential which is directly related to a gate voltage.
We determined numerically how the adiabatic increasing or diminishing of the impurity charge can be effectively accomplished by varying the chemical potential.  As we have shown, the static polarization
in a magnetic field strongly depends on the position of the chemical potential relative to the Landau levels. If the chemical potential is situated inside a Landau level, then the screening is very intense
and the effective charge of the impurity is strongly reduced. In addition, a nonmonotonic momentum dependence of the static polarization function with a peak at $q=0$ leads to oscillations of the screened potential with the sign change as a function of distance. On the other hand, if the chemical potential
lies between Landau levels, then the screening is minimal and the impurity can significantly affect the electron spectrum. These features of a charged impurity in graphene in the magnetic field are clearly observed in recent experiments \cite{Mao,states}.

\begin{acknowledgments}
This work is supported partially by the Program of Fundamental Research of the Physics and Astronomy Division of the NAS of Ukraine.
\end{acknowledgments}

\end{document}